# Experimental validation of the intensity refractometry principle for density measurements at the edge of a tokamak

M. Usoltseva[1,a)], S. Heuraux[2], H. Faugel[1], V. Bobkov[1], H. Fünfgelder[1], G. Grenfell[1], A. Herrmann[1], I. Khabibullin[3,4], B. Tal[1], D. Wagner[1], D. Wendler[1,5], F. Zeus[1] and ASDEX Upgrade Team[1]

[1]*Max-Planck-Institut für Plasmaphysik, Boltzmannstr. 2, 85748 Garching, Germany*
[2]*Université de Lorraine, CNRS Institut Jean Lamour, BP50840, F-54011 Nancy, France*
[3] *Universitäts-Sternwarte München, LMU, Scheinerstr. 1, 81679 Munich, Germany*
[4] *Max-Planck-Institut für Astrophysik, Karl-Schwarzschild-Str. 1, 85748 Garching, Germany*
[5] *Physik Department E28, TUM, 85748 Garching, Germany*

a)Author to whom correspondence should be addressed: maria.usoltseva@ipp.mpg.de.

Experimental validation is presented for a new type of microwave diagnostic, first introduced in the theoretical study in M. Usoltceva et al., Rev. Sci. Instrum. 93, 013502 (2022). A new term is adopted for this technique to highlight its difference from interferometry: intensity refractometry. The diagnostic allows measuring electron density, and in this work, it is applied at the edge of a tokamak. The implementation of this technique at ASDEX Upgrade, called Microwave Intensity refractometer in the Limiter Shadow (MILS), provides the first experimental proof of the diagnostic concept. Densities predicted by MILS are compared to several other diagnostics. The agreement and discrepancy in various radial regions of the density profile are analyzed and possible reasons are discussed. A wide density coverage is shown in the example discharges with densities from $2*10^{17}$ m$^{-3}$ to $2*10^{19}$ m$^{-3}$ at the limiter position. In these experiments, the radial location of the measurements varied from 5 cm in front of the limiter (up to 1 cm inside the separatrix was measured) to 3 cm in the limiter shadow. Experimental challenges of MILS operation and data processing are presented.

## I. INTRODUCTION

The density at the edge of a tokamak is a highly variable quantity. First, the diversity of plasma scenarios leads to differences of many orders of magnitudes in various experiments. In the example of ASDEX Upgrade (AUG) tokamak, density at the limiter can be as low as $10^{16}$ m$^{-3}$ in dedicated experiments[1] or can surpass $10^{19}$ m$^{-3}$ [2]. Second, the tokamak edge has one of the strongest spatial gradients of density in the entire plasma volume, so a variation of several orders of magnitude can take place within a few centimeters. Third, the fluctuation of edge density on short timescales can be comparable to or can exceed the time-averaged values, and therefore these variations would define the conditions for phenomena taking place in the edge plasma.

The edge conditions play an important role for many aspects of fusion plasma physics, such as the plasma core-edge mutual influence[3], the control of the exhaust and the first wall/divertor thermal loading[4], the properties of the Ion Cyclotron Range of Frequency (ICRF) heating in terms of power coupling[5] and of plasma-wave interaction at the edge[6], etc. One way to improve the understanding of the physics is to extend the ways to diagnose the plasma parameters.

In this work, we experimentally test a recently developed diagnostic applicable for tokamak edge density measurements, for which we are introducing a new term "intensity refractometry" (not to be confused with the existing time-of-flight refractometry[7] or any other refractometry technique). The principle of this microwave diagnostic is based on the refraction of the probing microwave beam, which is launched tangentially through the edge plasma (Fig. 1). From the measured change of the phase and the amplitude of the probing wave, relative to the case of propagation through vacuum, it is possible to reconstruct a radial density profile within several cm, by using the specially developed data interpretation methods based on the 3D full-wave synthetic diagnostic. While the numerical methods used for the data interpretation play crucial role in the diagnostic usage, this work focuses on the application of the already developed methods to experimental data, rather than on the examination of the numerical methods themselves. Details on the data processing techniques are not repeated in the main text of the article; they can be found in the theoretical study[8] and some discussion is provided in the Appendix. While this article describes the performance of the intensity refractometry in a tokamak, the diagnostic principle can be widely applicable to other plasma experiments, whenever suitable parameters can be found for the target conditions.

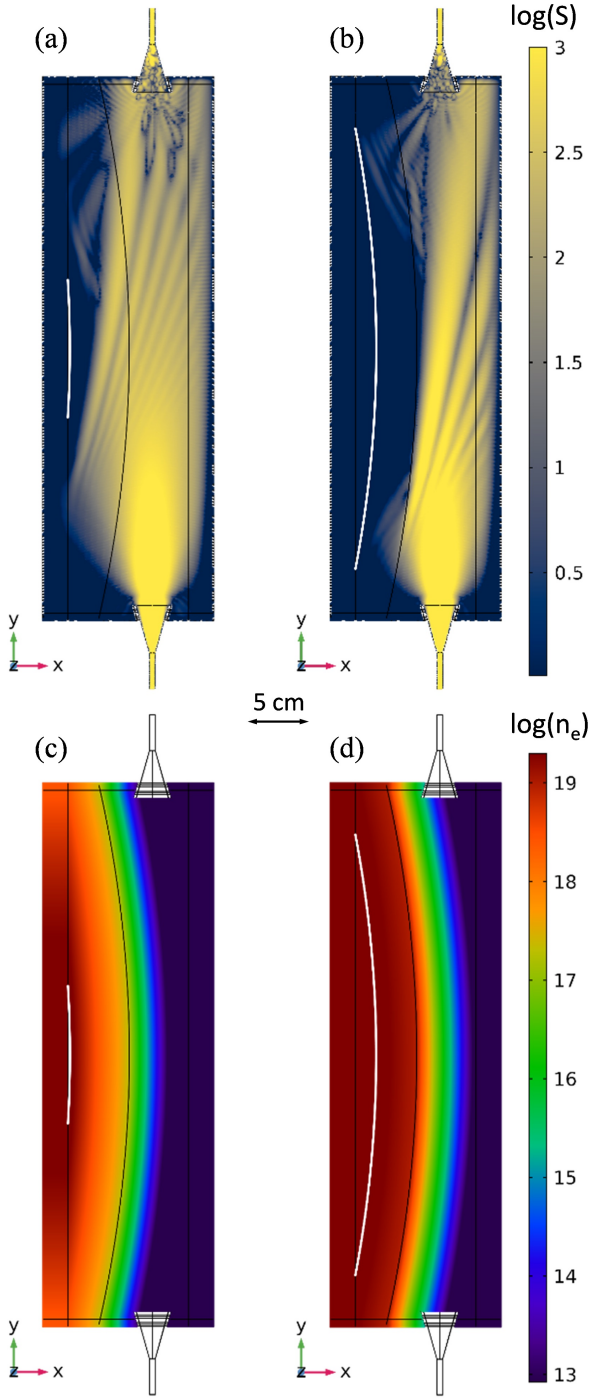

FIG. 1. Intensity refractometry principle: (a,b) refraction of the probing microwave beam from the higher density plasma on the left (Poynting vector in log scale plotted), (c,d) plasma density in log scale. The probing wave is launched from the bottom horn antenna with O-mode polarization. The magnetic field is directed along z axis. A radial-poloidal cross-section of a 3D model is shown. Densities, as defined by the analytical formula (see Section II.A) with the coefficients ($n_{lim}, a_{in}, a_{out}$), are: (a,c) (2e17,-100,-320), (b,d) (1e19,-30,-300). The limiter contour is shown with the black curve and the location of the probing wave cut-off layer is plotted in white.

The diagnostic implementation, installed in AUG and called Microwave Intensity refractometer in the Limiter Shadow (MILS) (called interferometer in the previous publications[8,9]), is targeted at covering the wide range of densities at the edge of AUG. The radial location of the measurements both in front of the limiter and in the limiter shadow covers the region of steepest density gradients; the estimation of the steepness of these gradients is one of the primary goals. Moreover, the diagnostic time resolution, defined in this study by the 0.2 MHz data acquisition rate, allows distinguishing density fluctuations on very short time scale.

In this paper, first comparisons of the outputs of the experimental and the synthetic MILS diagnostic are presented, which confirm the validity of the theoretical basis of the intensity refractometry technique. Comparisons of the MILS reconstructed density profiles with the data from other diagnostics are performed; they allow observing crucial points of agreement of the results, as well as investigating questions of partial discrepancies.

In the conditions of a fusion plasma experiment, various factors can disturb the measurements, such as stray magnetic radiation acting on the electronics, signal pollution by secondary sources at the same frequency, in-vessel components degradation due to the close contact with plasma, etc. We investigate possible sources of the data distortion and present approaches used to mitigate it.

## II. INTENSITY REFRACTOMETRY

### A. Principle

The intensity refractometry diagnostic technique[8] is based on launching a probing microwave beam tangentially through the edge plasma, in the O-mode polarization for density measurements. Parts of the probing beam can propagate directly, be partially deflected or be fully refracted away from the receiver (Fig. 1a,b). These processes depend on the antenna radiation pattern, probing wave frequency and the plasma density in the measurement volume. As can be seen in Fig.1, almost the whole probing beam is refracted earlier than it is able to reach the cut-off density layer (white curve), so no significant cut-off reflection occurs. A complex interference process of the probing beam parts takes place in the receiver antenna. At the end of the receiver waveguide, the wave power and phase are measured. In order to finely differentiate between various plasma conditions, the frequency and the geometrical parameters of the intensity refractometer (axis length, antenna field pattern, antennas rotation and shift relative to each other) need to be balanced, such as to achieve a large variation of the measured phase shift and power change.

To reconstruct the density radial profile from the measured quantities, we utilize the database of about 2500 cases, modelled in the MILS synthetic diagnostic[8]. In the database, all profiles have the shape of an exponentially decaying function of the distance $x$ to the limiter $n_e = n_{lim} * exp(a_* * x)$ (where $n_{lim}$ is the density at the limiter and $a_*$ is the inverse decay length), with a bend at the limiter location, as shown in Fig. 2. Such shape can be typically seen in the experimental data[10] and the exponentially decaying profiles are widely used in models as an approximation for the tokamak edge region[3]. The much stronger decay in the limiter shadow is expected because of the large decrease in the connection length after the limiter[11]. While the transition should happen smoothly, we take a simple bend at one point in our model, to be able to describe the profile by three variables ($n_{lim}, a_{in}, a_{out}$) (where $a_{in}$ is the inverse decay length in front of the limiter and $a_{out}$ is the inverse decay length in the limiter shadow). To obtain density values from the MILS experimental measurements, modelling cases with the closest output values are found, weighted and a resulting density profile is calculated by using the dedicated genMILS method[8]. The error in density is directly related to the errors in the phase and power. The total error contains the contributions of the numerical error, the genMILS method error and the experimental error. Besides the previous theoretical work[8], some discussion of the MILS density reconstruction procedure can be found in the Appendix.

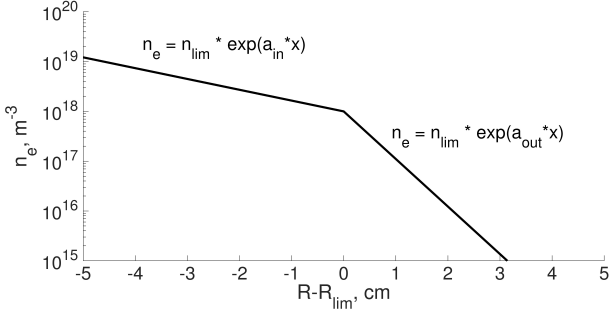

FIG. 2. Shape of the density radial profile use in the MILS synthetic diagnostic The profile is defined by three variables (n_lim, a_in, a_out).

It is useful to demonstrate the inapplicability of the classical interferometry approach to the analysis of the measurements of MILS or intensity refractometry in general. The phase shift $\Delta \varphi$, measured by an interferometer, is proportional to the integral density along the path between the emitter and the receiver[12]:

$$\Delta \varphi \approx \frac{\pi}{\lambda_0 n_c} \int_0^L n_e(x) dx \quad (2)$$

where $\lambda_0$ is vacuum wavelengths, $n_c$ is the cut-off density, $n_e(x)$ is the density as a function of the coordinate $x$ along the interferometer axis and $L$ is the interferometer axis length.

If we calculate this value for all density profiles of the MILS synthetic diagnostic database and compare it to the actual phase shift measured by the synthetic diagnostic, we can see that these two values are very different for almost the whole database (Fig. 3). As expected, the part of the probing beam, which passes directly from the emitter to the receiver, constitutes only a small contribution to the whole wave interference pattern, which reaches the receiver. The interpretation of MILS measurements by using interferometry formula would provide misleading results.

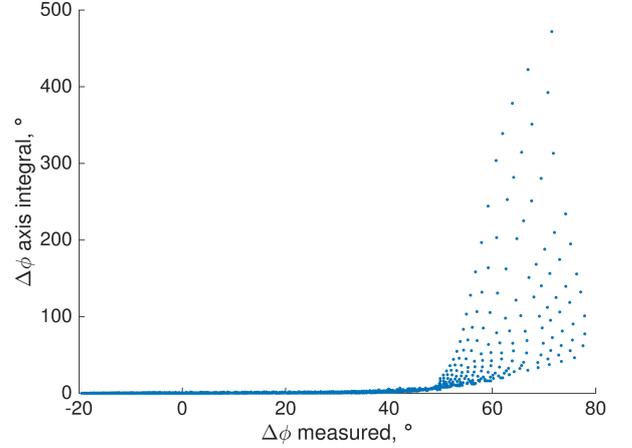

FIG. 3. Comparison of the phase shift measured by the MILS synthetic diagnostic and the phase shift calculated analytically by taking into account only the density along the MILS axis, corresponding to the interferometer measurement principle.

## B. Application in ASDEX Upgrade

The implementation of the intensity refractometry technique in AUG is shown in Fig. 4a. The distance between the emitter and the receiver antennas is 429.1 mm. The horn antennas are identical and have the radiation pattern with the 3 dB beamwidth of 17.2° in both planes. The emitter and the receiver are aligned on one line, in the direction tangential to the limiter contour and are located symmetrically relative to the limiter contour. The probing wave is sent at 47 GHz and the antennas are accurately aligned with the short edge of the waveguide parallel to the total (toroidal plus poloidal) magnetic field to provide the O-mode propagation.

The choice of the parameters was guided by the available space and the AUG typical plasma shape and size, as well as by the expected densities and the need to obtain large phase and power variation of the probing wave for a broad density range. The MILS axis needs to be long enough and/or the frequency has to be low enough to obtain significant phase shift for low densities. At the same time, too big axis length or too low frequency should not be taken to avoid phase fringe jumps and loss of the signal due to too strong refraction in the high-density plasma scenarios. Deeper probing inside plasma is achieved by placing horn antennas as close as possible to the plasma, although too close proximity should be avoided, as it could lead to the antenna damage or plasma contamination by the sputtered material. The radiation pattern of the horn antennas should also correspond to the objective of detecting the probing wave in a large range of densities without losing it at the receiver position. The previous theoretical study[8] of the intensity refractometry technique was performed for the configuration described in the current work. It was found that the chosen MILS geometrical parameters, frequency and position relative to plasma result in a wide coverage of the density span and high accuracy of the density reconstruction in the range of interest (the accuracy decreases notably for very low densities, in the range which is only rarely present in experiments).

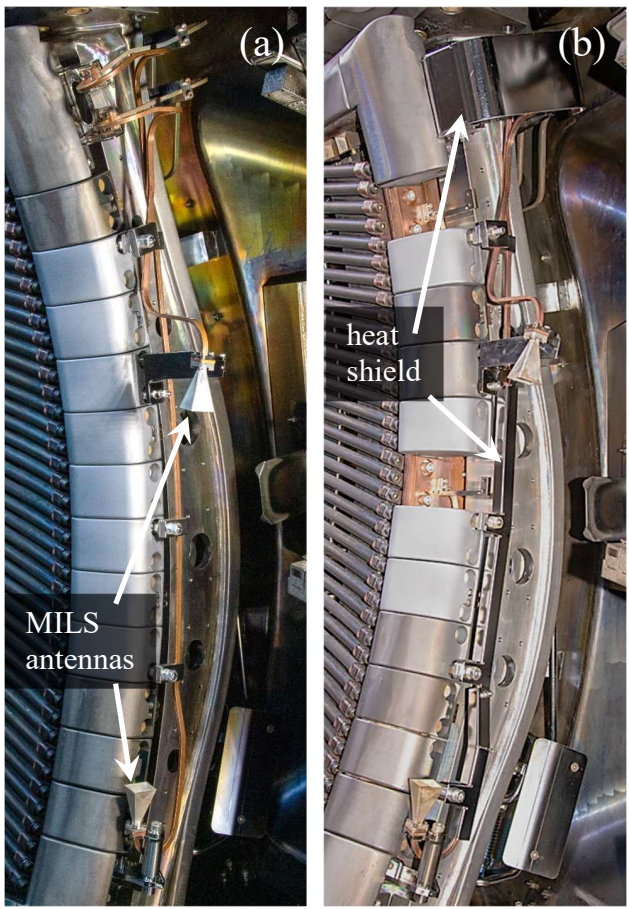

FIG. 4. MILS diagnostic in AUG with two horn antennas and microwave waveguides: (a) without the thermal shield, (b) with the thermal shield.

Heat loads might lead to thermal expansion and movement of in-vessel components in a tokamak. It is ensured that the MILS antennas stay precisely in their position and can neither shift nor rotate, by fixing their support on the cooling frame of the ICRF antenna, which experiences no thermal effects due to active cooling. Precise measurements (with 0.1 mm accuracy) of the horn positions before and after the experimental campaign have confirmed that no shifts occurred.

Presence of objects in the probing area of MILS, which are not transparent for microwaves, could lead to the measurements being affected by reflections on these objects. All such objects have to be added in the simulation domain of the synthetic diagnostic, to make sure that their influence on the probing wave propagation is taken into account in the data interpretation. In the current implementation, the 3 dB beamwidth of 17.2° results in the probing beam width of 32 mm in the middle of the MILS axis length. The object closest to MILS inside the AUG vessel is the limiter of the ICRF antenna, with the distance of about 5-10 cm to the center of the MILS axis. It is far enough to ensure that the main lobe of the antenna radiation pattern is not affected by any reflections from the ICRF antenna limiter. In addition, the surface of the limiter is not parallel to the MILS axis, which makes it unlikely that any significant power can be reflected even from the side lobes of the MILS emitter radiation. In other implementations of the intensity refractometry technique, spurious reflections should be always addressed with care and taken into account in simulations, if present.

The electronics scheme of the MILS system components is shown in Fig. 5. We use two Kuhne MKU 47 G2 transverter modules, one configured as a transmitter, the other one as a receiver (marked as "f × 4" in the scheme). The reference oscillator Kuhne MKU LO 8-13 PLL provides the signal at 11.736 GHz and 20 mW, which is then multiplied by four in the transverter modules, giving 46.944 GHz. Furthermore, an intermediate frequency source is required, for the upper sideband image rejection mixer of the transmitter and the receiver, resulting in the operating frequency of 47.089 GHz. Since the receiver and the transmitter use the same reference signal, split by a power splitter, and the local oscillator has two synchronized output signals, the whole system is coherent and permits phase and power detection at 145 MHz.

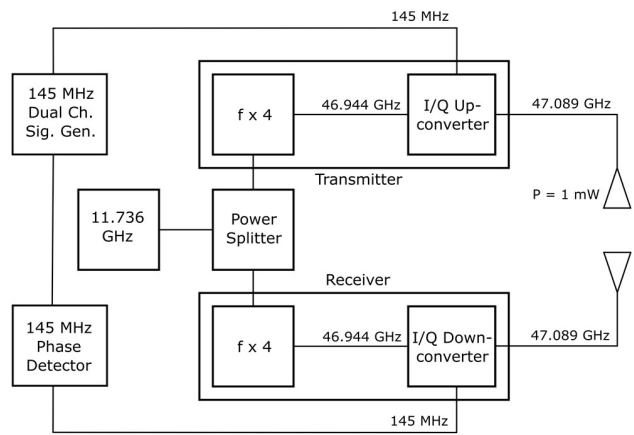

FIG. 5. Scheme of the MILS electronics.

The measurements are done with a detector that uses two Analog Devices AD8302 phase detector integrated circuits at the intermediate frequency of 145 MHz. From the dual channel signal generator Siglent SDG 6022X, one signal is sent through the "plasma path", i.e. it goes through the described above electronics and crosses the plasma between the two horn antennas, and the other signal goes directly to the 145 MHz detector. The second signal is needed, because it gives a reference value for the detection of the phase change in plasma, in the same way as it is done in the interferometry. In the detector, the power of the waves in both channels is measured (in logarithmic scale), as well as the phase difference between the two waves. A phase shift of 90° is done for one of the signal paths, providing the possibility to detect the phase difference not only within 180°, but for the whole period of 360° (Fig. 6a).

The power of the signal, sent to the "plasma path", is typically around 3 mW. Considering the combination of signal attenuation in the cables and waveguides and signal amplification in the transmitter and the receiver, the power level of the probing wave reaching the plasma is about 1 mW. Very little power is reflected in the transmitter channel both in experiments and in the simulation ($\leq 0.1$ %). The signal decay or amplification in plasma can vary significantly depending on the density. Therefore, the power level reaching the detector changes within a large range. The lower limit of the power measurement of the detector is $10^{-7}$ W. The power of the reference path signal is set to be equal to the vacuum power level of the "plasma path". Typically, this value is 0.03 mW = -15 dBm, corresponding to 1.6 V on the detector power channels calibration curve (Fig. 6b).

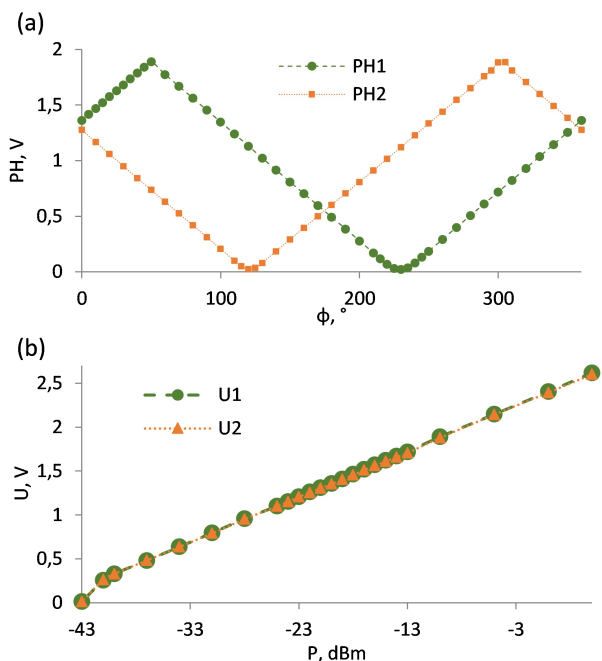

FIG. 6. Calibration curves for the detector outputs: (a) phase φ, with two outputs PH1 and PH2 providing the possibility to measure a phase shift of up to 360°, (b) power P for two channels.

## III. COMPARISON OF DATA FROM EXPERIMENTAL AND SYNTHETIC MILS DIAGNOSTIC

The database of all modelled cases is shown in Fig. 7. In the same figure, experimental data measured by MILS in several plasma discharges are plotted. Mean signals are shown, for which the moving mean is calculated in the 50 ms window, ensuring good filtering of the signal oscillation due to density perturbations. Such large window is chosen to focus on the well-averaged signals only, since the current comparison is not aimed at the detailed characterization of the signal perturbation amplitudes and frequencies in each of the chosen shots. The 50 ms window taken for this analysis should not be regarded as the limit of the MILS temporal resolution. Plasma conditions for this comparison were chosen such that there are no large asymmetric oscillations in the MILS signal and the plasma is quite well aligned with the limiter contour, as in the model. The shots 37973 and 39023 were performed specifically to obtain large variation of density in the plasma edge and to compare measurements from different diagnostics (see Section IV). Other discharges with some density variations were also found in the AUG database.

The experimental data is well constrained within the the boundaries of the modelling database. Such good agreement would not be achieved, if the shape of the density profile, chosen for the modelling, would be too far from the experimental density profiles. Therefore, this agreement supports the suggested possibility to approximate the experimental profiles closely enough by the chosen simplified shape.

In comparison, strong deviations from the modelling database are observed when large perturbations are present in the field of view of MILS[9]. Examples of it are local density modification when ICRF is used or ELMy plasma with large filaments at the edge. The MILS data collected in such conditions cannot be analyzed by the standard procedures and requires dedicated methods. For the case of ELM filaments, both experimental and synthetic MILS diagnostic show[9] that the measured values go out of the database and form similar "trajectories" outside of the database on the phase-power diagram. Dedicated methods are used to analyze these "trajectories", in order to obtain filament properties. For ICRF-induced density modifications, a special analysis routine is being developed. The database presented in this paper is constructed for monotonic profiles with constant densities along the flux surfaces and therefore should not account for profiles with large density perturbations.

## IV. COMPARISON OF DENSITY PROFILES TO OTHER DIAGNOSTICS

### A. Challenges in comparisons between diagnostics

A comparison between diagnostics, located at different toroidal and poloidal positions in a tokamak and not magnetically connected to each other, is not straightforward. When density profiles are recalculated to the normalized radius coordinate $\rho$, it is expected that the values from different diagnostics would match at the same normalized radius (at the same flux surface). While this assumption is typically fulfilled in the region in front of the limiter, in the limiter shadow the local conditions can potentially cause large differences, so the validity of this assumption remains an open question and should depend on the conditions of the plasma. This aspect is important in the present study, since the MILS measurement region lies partially in the limiter shadow.

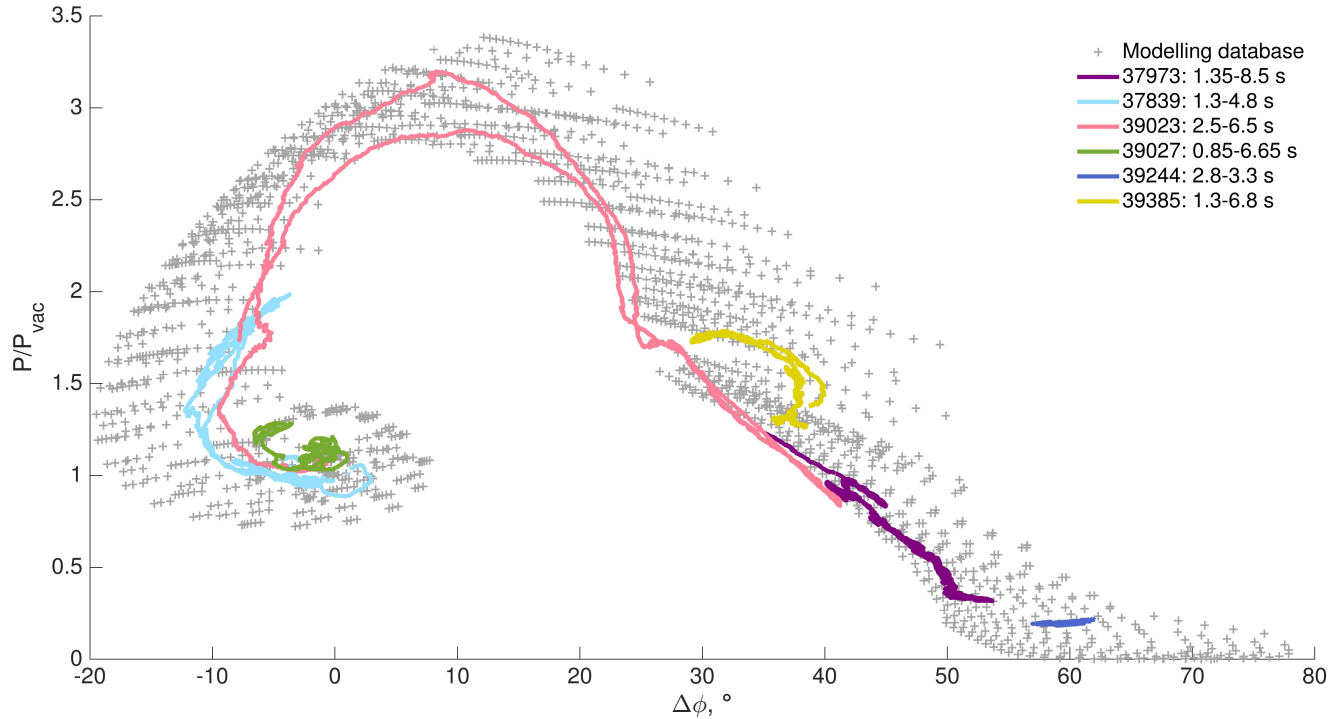

FIG. 7. MILS modelling database and experimental data from several shots. The experimental values are averaged, with signal oscillation removed.

Additionally, the location of the limiter might be not well defined. Only in an ideal case of plasma shape perfectly matching the whole limiter curvature, it would be possible to say that there is a certain $\rho$ value, where the limiter shadow region starts. In a general case, when the alignment is not ideal, the most protruding point of the limiter can be taken as the lowest $\rho$ of the limiter shadow, a so-called "global" limiter. In this case, not only the local conditions might play a role for each diagnostic, but also the position of its "local" limiter relative to the "global" one.

Another factor, which can affect the comparison of measurements of different diagnostics, is their temporal resolution. Density fluctuations of significant amplitude, when not correctly resolved and processed in the signal, can lead to erroneous determination of the mean background signal and inability to separate the background and the fluctuation contributions. Since the diagnostics used in this study have very different time resolution, the best strategy for the comparison is to choose data from plasma discharges with as low as possible level of density perturbations. Therefore, plasma scenarios are taken, which have small-amplitude filamentary density fluctuations in the SOL plasma. Still, some influence of these fluctuations on the evaluated averaged density can remain and contribute to discrepancies between diagnostics.

### B. Employed diagnostics and their parameters

The diagnostics used in this study for the electron density comparison are the He-[13] and Li-Beam[14] Emission spectroscopy (BES) and Langmuir probes (LP) on a movable manipulator[15]. All of the diagnostics, including MILS as well, are located at different toroidal and poloidal positions in AUG and are not directly magnetically connected. Each diagnostic has its own boundaries for the radial range of measurements and its own lower limits of density detection.

The He-BES is targeted more at the near SOL, with the last observed point at about 1.5 cm in front of the limiter. There is no strict lower limit for the density, but the data with not high enough amplitude during the active phase of the gas puff, relative to the phase with the gas puff turned off, are discarded. Such values were all below $10^{18}$ m$^{-3}$ in the presented data. For the Li-BES, the outermost channel is 2 cm in the limiter shadow and the lower limit of density measurement can be taken as $5*10^{17}$ m$^{-3}$. The LP were moved radially by few cm in the limiter shadow in the shot 37973, and in the shot 39023 they stayed at a constant position of about 0.5 cm in the limiter shadow. Some LP data was excluded, for which the signal-to-noise ratio in the raw data was too low or the calculated temperature was out of realistic bounds. In the shot 37973, such points were found and cut beyond 3 cm in the limiter shadow. No specific density lower limit was applied.

The He-BES and the Li-BES data for this study were taken with 1 and 10 ms time resolution, respectively (time needed to collect enough light for good signal-to-noise ratio). For LP, one voltage sweep takes 0.5 ms and defines the temporal resolution of the density evaluation. A radial density profile is obtained by the manipulator movement during 100 ms. For MILS, the time resolution is 5 µs in the presented measurements. To perform a uniform comparison, for all diagnostics a single density profile is calculated, averaged over 100 ms (for LP all points during 100 ms are taken).

The errors in the LP data in the shot 37973 are calculated using data from a probe constantly biased at negative voltage. The relative fluctuation of the ion saturation current from this probe is sampled at 2 MHz and provides an estimate of the expected variations, which are not resolved by the probe with the swept voltage. The obtained relative fluctuation value of 60 % exceeds the error, which comes from the density evaluation method from the swept Langmuir probe measurements. Therefore, the value of 60 % is taken for the error bars in the shot 37973. In 39023, the dispersion of the plotted data captures its statistical variation during 100 ms. For the error bars of Li-BES and He-BES data, the largest value is taken from the two options, their variation in the chosen time period and the error of the density calculation method. For the radial profile measurements, the manipulator, where the probe is installed, is constantly moving. The obtained measurements, shown as distinct radial points, represent signal collected over a radial range. The corresponding error should be small compared to the taken total error, so it is not accounted for in the current study.

For MILS, the errors in power and phase from the modelling[8] ($\delta\varphi_{mod} = 0.5°$ and $\delta P_{mod} = 1.5\,\%$) are summed up with the experimental errors ($\delta\varphi_{exp} = 0.3 - 0.6°$ and $\delta P_{exp} = 0.9 - 1.2\,\%$ for all examples except 37973, 7.96-8.07 s, for which $\delta\varphi_{exp} = 12.9°$ and $\delta P_{exp} = 7.9\,\%$ are much larger because of large radiated energy causing MILS signal drift, see Section V.G); the sum defines the error bars of the density values. In the considered experiments, the shape of the flux surfaces is quite well aligned with the limiter contour; a possible error due to misalignment has not been quantitatively investigated and this analysis is planned in the future work. The error bars do not account for possible profile shape deviation from the double-exponential profile shape, used for the database construction. While the density values and the error bars for MILS are calculated at every 0.5 cm in the genMILS algorithm and the points are treated as independent in the density reconstruction, these values are not independent[8]. The possibility to obtain radially varying level of error comes from the fact that some parts of the density profile make larger influence on the MILS measured values and they can be predicted with smaller uncertainty, as the study of the MILS density reconstruction shows[8] (see also Appendix for more details).

### C. Comparison results

The density comparisons are done in two of the shots, shown in Fig. 7, 37973 and 39023, in which dedicated density scans were performed. The two shots together cover well the parameter space and are representative for the MILS measurements. In Fig. 8 and 9, density profiles from all employed diagnostics are shown for several time slots in the two discharges.

The agreement of the density values between various diagnostics is very good in the region in front of the limiter. The values from different diagnostics overlap within one error bar range. In the limiter shadow, the results are less uniform. A large part of data in the limiter shadow coincides within the one error bar range or at least within the three error bar range. This is the case for all comparisons of MILS and LP data. Li-BES and MILS points, which lie outside the three error bar intervals, disagree between each other in the high-density examples in Fig. 9. In the high-density conditions of the discharge 37973, it is observed that MILS tends to predict steeper density gradients in the limiter shadow than Li-BES, and the LP data lies in between MILS and Li-BES estimations. For lower densities, measurements of MILS and LP coincide well, but there is not so much data from the other diagnostics for comparisons.

Since the agreement between various diagnostics is notably better in front of the limiter than in the limiter shadow, the main reason for larger variations in the limiter shadow can be different local conditions near different diagnostics. Other factors, described in the part A of this section, can also contribute to discrepancies. For MILS, the location in AUG was chosen at the side of an ICRF antenna, since one of its applications is to study local density modifications introduced by ICRF. Even if each diagnostic is sensitive to its own surroundings and provides different measurements in the limiter shadow relevant to a certain location, MILS measurements at its position can provide data for the local conditions in the limiter shadow, useful for ICRF studies (with dedicated data analysis approach).

The plots from the beginning of the discharge 39023 highlight particularly well the importance of the MILS measurements in the almost non-diagnosed region of the far SOL plasma. In the absence of the LP data, which was obtained in this shot with special settings at the location in the limiter shadow and which is not routinely available in AUG experiments, there would not be any source of density values around the limiter radial position. A very large range of densities is accessible for MILS, which can be seen from the given examples, with densities from $2*10^{17}$ m$^{-3}$ to $2*10^{19}$ m$^{-3}$ at the limiter position. The radial range of MILS measurements changes depending on the density, as can be seen in the given examples (because of different refraction conditions for the probing wave). Density at 1 cm inside the separatrix was measured at 3.4-3.5 s in the shot 39023, due to a small plasma-wall gap. In the higher density scenarios, only the part further away, from the limiter towards the wall, is diagnosed by MILS, as was also demonstrated in the theoretical study[8].

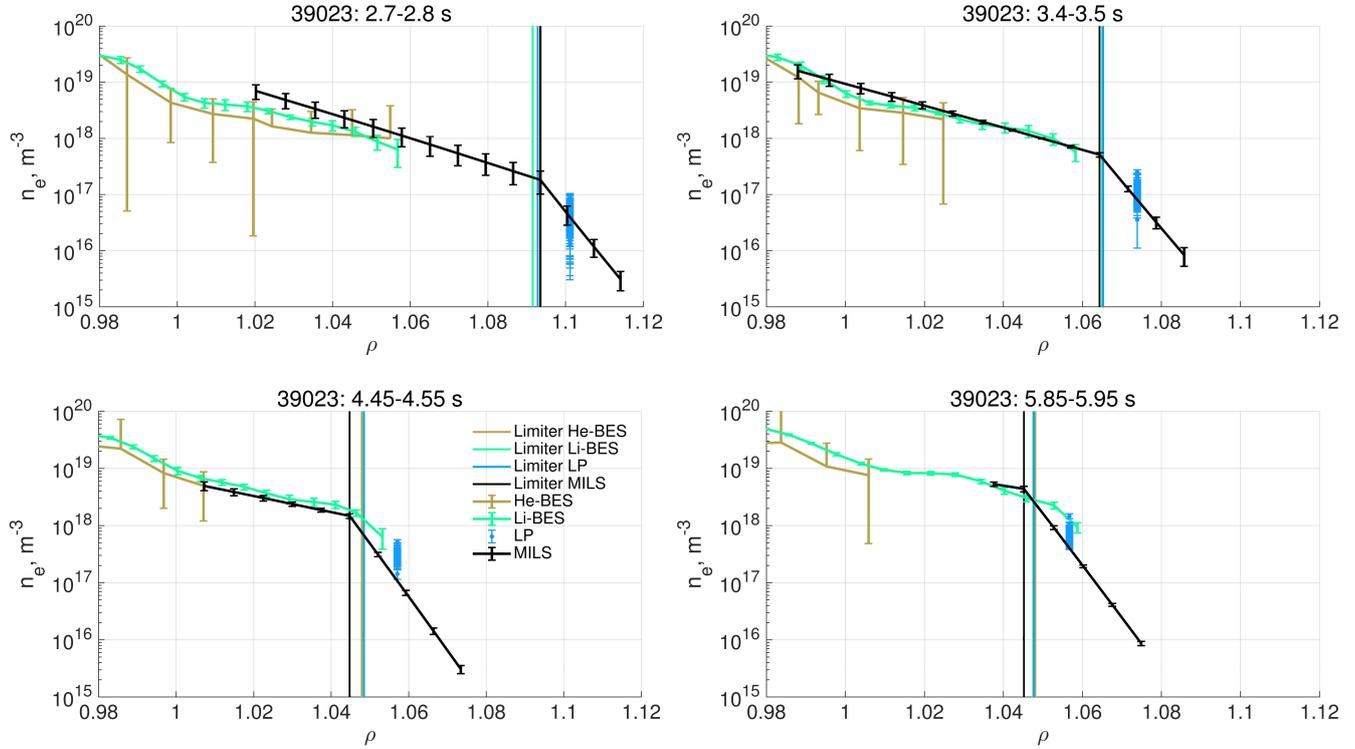

FIG. 8. Comparison of edge density profiles between several diagnostics for the shot 39023 in normalized radius $\rho$ coordinates.

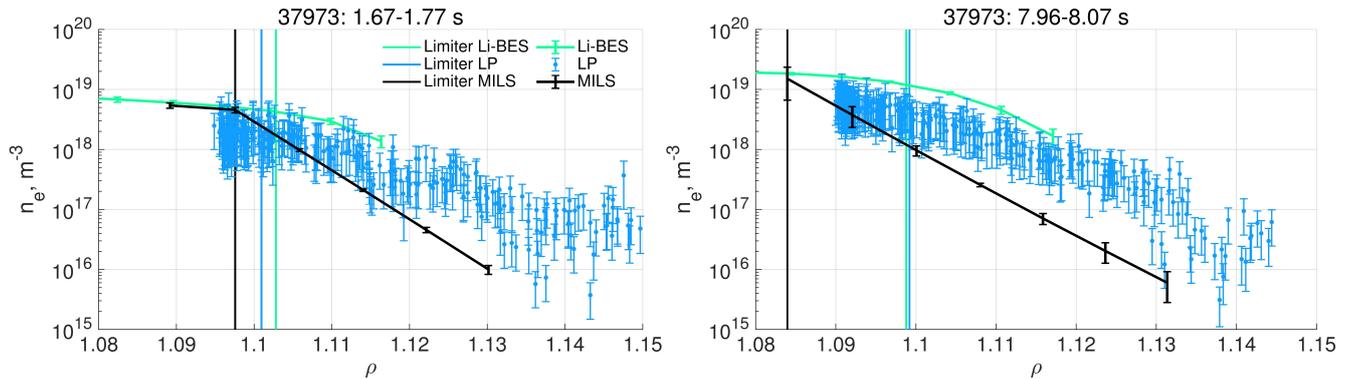

FIG. 9. Comparison of edge density profiles between several diagnostics for the shot 37973 in normalized radius $\rho$ coordinates. He-BES values are not plotted, as they are out of the $\rho$ range.

### D. MILS output with density from another diagnostic

A procedure opposite to the profile reconstruction for MILS can be done, when an already known density profile is used in the MILS synthetic diagnostic as an input. For example, it can be tested whether MILS provides a similar output when a density profile from another diagnostic is used. In the studied cases, the shapes of the profiles from other diagnostics (where available) are mostly quite similar to the profile shape used by MILS, decaying exponentially until the limiter and then having a steeper but still constant exponential slope after the limiter. The exception from this can be seen for the Li-BES in the shot 37973, where the decay length only changes slowly around the limiter and keeps decreasing further towards the wall. The differences between the profiles are bigger in the shot 37973, so this discharge is more interesting to examine.

It was decided to first test a profile, which approximates closely the Langmuir probes (LP) data at one time interval in the middle of the shot 37973 (see Fig. 10). The density values of the chosen test profile are in between the MILS and the Li-BES densities. The test profile is used as an input in the MILS synthetic diagnostic, and the obtained value of ($\Delta\varphi, P/P_{vac}$) is plotted in Fig. 11. The difference between the value measured by MILS in the experiment at the same time slice (also plotted) and the output for the test profile is enormous and cannot be explained by any uncertainties in the MILS measurements. The total MILS errors for the measured values amount to $\delta\varphi = 6.8°$ and $\delta P = 5.2$ %, as shown by the error bars (quite high in this particular shot, see Section V. B and V.G for details). The MILS signal fluctuations in time over the 0.1 s window are plotted in Fig. 11 and are smaller for the power and much smaller for the phase compared to the errors, so they also cannot explain the obtained result. Therefore, the observed differences in the profiles should be explained by other reasons, including the differences in the local conditions near each diagnostic discussed in the Section IV.A.

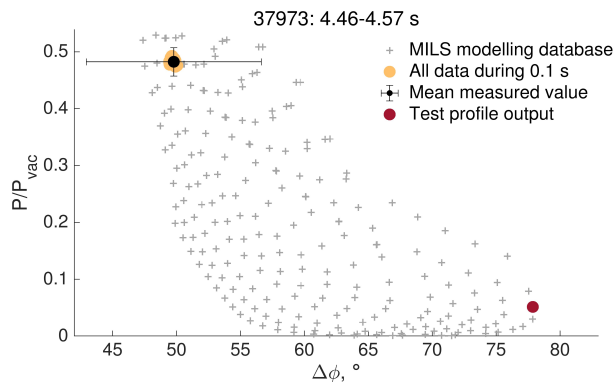

FIG. 11. Comparison of the MILS experimental measurements (with all data points during 0.1 s, the mean value and the errors indicated) and the calculated synthetic diagnostic output for the test profile from Fig. 10.

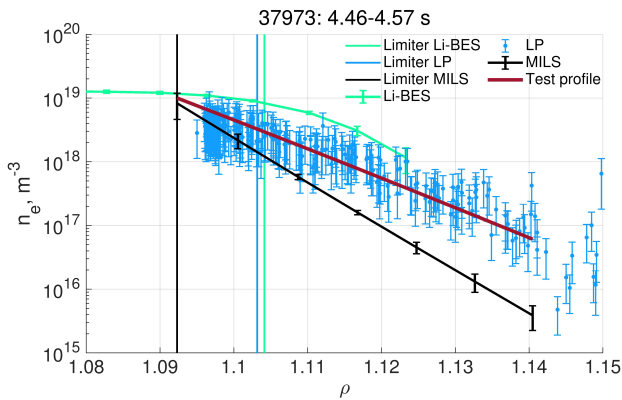

FIG. 10. Density profiles from several diagnostics and a test profile to approximate LP data, which is used as a MILS synthetic diagnostic input.

This analysis gives an example of how the error level in the MILS measured values, phase and power, is connected to the errors in the MILS reconstructed density profile. The values of the power and phase for the test profile lie far out of the error boundaries of the MILS measured values, in the same way, as the density values of the test profile are quite far from the range of density given by the MILS reconstructed profile plus the error bars.

The test was not continued with the profile, which approximates the Li-BES profile, or with any other larger density values, since the point obtained in the first test already lies on the boundary of the MILS database. The much higher densities, measured by the Li-BES, would provide a result, which is only further away from the MILS measured value.

## V. EXPERIMENTAL FACTORS INFLUENCING MILS DATA

We have identified some experimental factors, which can distort MILS signals, as well as analyzed other factors, which could potentially affect the measurements, but do not seem to make any influence. The overview is given below.

### A. Stray magnetic field

The attenuation of the WR19 (U-band) waveguides, used to transfer MILS signal from inside the tokamak to the electronics, is about 15 dB for approximately 10 m of the waveguides[16]. In order to avoid larger power loss, the waveguides do not extend further, so the detectors are placed relatively close to the tokamak, which means close to the magnetic field coils. The stray magnetic field can reach tens of mT around the electronics rack.

Most of the MILS electronic components are not sensitive to the magnetic field influence, but some of them were observed to switch off during a plasma shot. If the current in the ASDEX Upgrade V3 coils (two of the whole set of twelve vertical field coils), nearest to the electronics rack, exceeded a certain level, it caused a loss of the microwave signal. The reason was that the DC-DC converters in the Kuhne MKU 47 G2 transverter modules tripped when the stray magnet field was too high. The solution, implemented by the manufacturer, was to bypass the DC-DC converters and change the supply voltage to 5.2 V instead of 12-14 V. In the new configuration, the transverter modules never showed the same failure. Another part of the electronics scheme, which also has DC-DC converters, is the 145 MHz signal generator. It has worked reliably during the MILS operation in the AUG experimental campaigns 2019/20 and 2020/21. Only when it was moved to another place within the electronics rack for some tests, it appeared that it could also be switched off for a few seconds, when the stray magnetic field reaches its highest possible values.

## B. Thermal expansion of the in-vessel parts

The microwave waveguides and horn antennas inside the AUG are located very close to the plasma, with the front edge of the horn antennas being 2.5 cm away from the limiter edge. Therefore, the in-vessel microwave components are exposed to both the plasma radiation and the hot particle fluxes. As a result, thermal expansion of the waveguides is expected, which can lead to drifts in the measured phase and power values. In addition, a drift in the measured signals might be caused by the movement of the emitter and receiver horns relative to each other, if the structures, which the microwave components are mounted to, can expand because of the heat loads. In our case, such relative displacement was excluded by attaching the supports of the waveguides to the cooling frame of the ICRF antenna limiter, which is kept constantly at low temperature. Therefore, the effects described below are attributed to the thermal expansion of the waveguides only.

### 1. Phase signal drift

As observed from the MILS data, the drift of the phase signal is well correlated with the plasma radiation, which is the dominant cause of the thermal expansion of the waveguides (Fig. 12). Infrared measurements confirm the correlation of the horn antenna temperature with the total radiated energy during a plasma discharge.

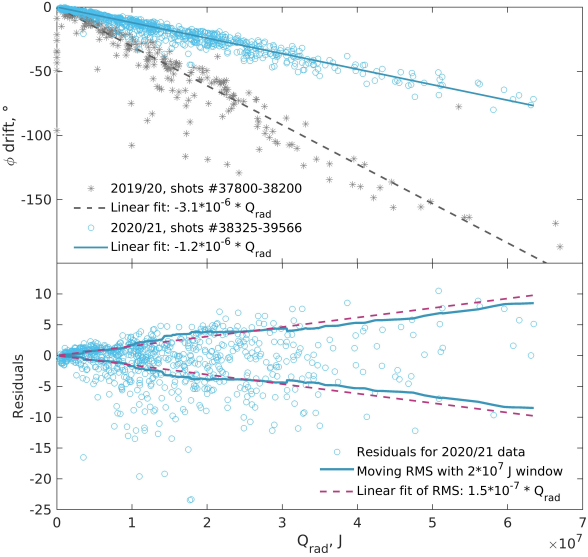

FIG. 12. Drift of the phase as a function of the radiated energy with a linear fit (top); residuals of the fit with the RMS (and RMS multiplied by -1) and linear fit of the RMS (bottom). The phase drift is shown for two experimental campaigns, before and after the thermal shield installation, and the residuals are plotted for the second campaigns.

In order to reduce the data distortion because of the thermal expansion, a cover was designed and installed between the 2019/20 and 2020/21 experimental campaigns, which shields around 80 % of the waveguides exposed to plasma (Fig. 4b). A reduction of the phase drift by the factor of 2.6 was achieved, as can be seen in Fig. 12, although it was not possible to eliminate the presence of this drift.

The phase drift leads to values shifted during a plasma discharge, as shown above, but it also causes the starting value at the beginning of the shot to vary during the campaign. However, since the data interpretation is based on the relative change of the phase, the shift of the absolute value from shot to shot does not introduce any error. Within one shot, a correction is done to compensate for the drift caused by the thermal expansion. It was observed that in a shot with relatively constant plasma parameters and radiation level, the phase drift is well approximated by a linear function in time (Fig. 13), due to the constant level of the radiation. Therefore, the correction for the phase drift $\varphi_{cor}(t)$ is done by adding a shift at each time point, proportional to the total phase drift in the discharge $\varphi_{drift}$ multiplied by the cumulative integral radiated power at that time point $Q_{cum}(t)$, normalized by the total radiated energy during the whole discharge $Q_{rad}$:

$$\varphi_{cor}(t) = \frac{\varphi_{drift} * Q_{cum}(t)}{Q_{rad}} \quad (2)$$

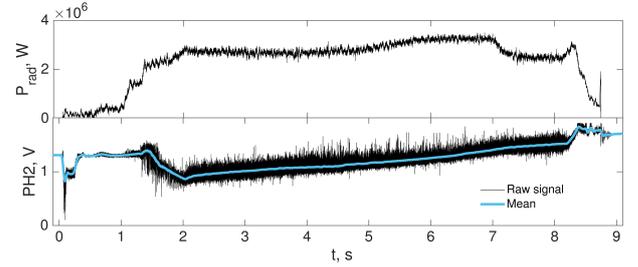

FIG. 13. Example of a plasma discharge with almost constant conditions (main plasma from 2 to 8 s). The drift in the phase is close to a linear function in time (one of the two raw signals PH1 and PH2 is shown, linearly proportional to the actual phase shift).

Since the correction might not represent the real phase drift precisely, it introduces an error, which was estimated from the RMS of the residuals of the linear fit of the phase drift and the radiated energy correlation (Fig. 12). The error is taken as a function of $Q_{rad}$, obtained by a linear fit of the RMS of the residuals, calculated in the moving window of $2 * 10^7$ J: $\delta\varphi_{drift} = 5.2 * 10^{-7} * Q_{rad} = 0.17\ \varphi_{drift}$ (for the AUG 2019/20 campaign) and $\delta\varphi_{drift} = 1.5 * 10^{-7} * Q_{rad} = 0.125\ \varphi_{drift}$ (for the AUG 2020/21 campaign). In the majority of the shots, the radiated energy is up to $2 * 10^7$ J, which corresponds to an error in phase of up to 10° in the 2019/20 experiments and up to 3° in the 2020/21 shots.

### 2. Power signal drift

In case of the power signal drift, the correlation with the plasma radiation is not so clear. While in the first experimental campaign the power signal drift seems to be caused by several factors, among which the radiation is one of the largest ones, the influence of the radiation became much weaker after the thermal shield installation (Fig. 14).

The same analysis is performed as for the correlation of the phase drift and the total radiated energy. The linear fit of the data shows a reduction of the power drift by the factor of 2.3 after the installation of the thermal shield, which is similar to the level of the improvement in the phase drift. If the correction is done in the same way as for the phase drift and the error is estimated likewise, it results in the errors: $\delta P_{drift} = 2.4 * 10^{-7} * Q_{rad}$ (for the AUG 2019/20 campaign) and $\delta P_{drift} = 1.2 * 10^{-7} * Q_{rad}$ (for the AUG 2020/21 campaign). In the shots with the radiated energy of up to $2 * 10^7$ J, the error in the power is up to 5 % in the 2019/20 data and up to 2.5 % in the 2020/21 discharges.

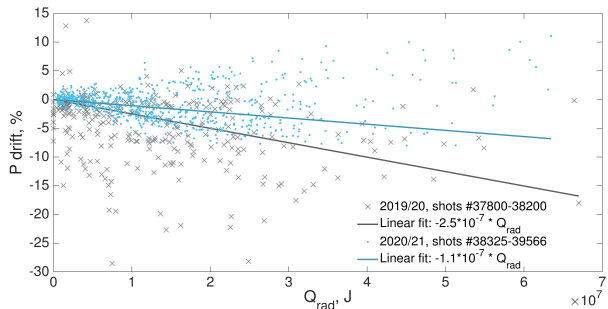

FIG. 14. Drift of the power as a function of the radiated energy with a linear fit, for two experimental campaigns.

## C. Deposition of a metallic layer in the waveguide

One of the proposed explanations of the power drift during a plasma discharge is the formation of a thin metallic layer inside the waveguide connection to the lower horn antenna. At that location, a Kapton (polyimide) foil is inserted in order to prevent dust deposition inside the waveguide. The foil covers the 2.388×4.775 mm² area of the waveguide cross-section, but it has a circular hole in the center of 1.55 mm diameter to avoid closed volume inside the waveguide, which can cause outgassing during tokamak operation. Between the experimental campaigns, it was noticed that the Kapton foil changed the color and a deposition of a metallic layer is suspected. The power values drift both in the negative and positive direction could be explained by the modulation of the transmission introduced by varying thickness of the metallic layer. In addition to the metallic layer, dust or even larger debris could fall on the Kapton foil and be shaken away from it from time to time, mostly during disruptions, leading to changes in the transmitted power.

We have performed a modelling with a mode matching code[17] in order to estimate the power change, which could be caused by such a metallic layer inside a waveguide. The scheme is shown in Fig. 15a. In the model, it is not possible to have a hole in the Kapton foil, instead the foil covers completely the waveguide cross-section and the metallic layer has a rectangular shape with a rectangular hole. Foils with the permittivity of 3.43 and with the thicknesses of 25.4 or 50.8 μm are used in the experiment. Both thicknesses options were checked in the simulation, but it had only minor impact on the result, so the results are shown for 25.4 μm. The metallic layer thickness $d$ is scanned within a range of 0.1–100 nm.

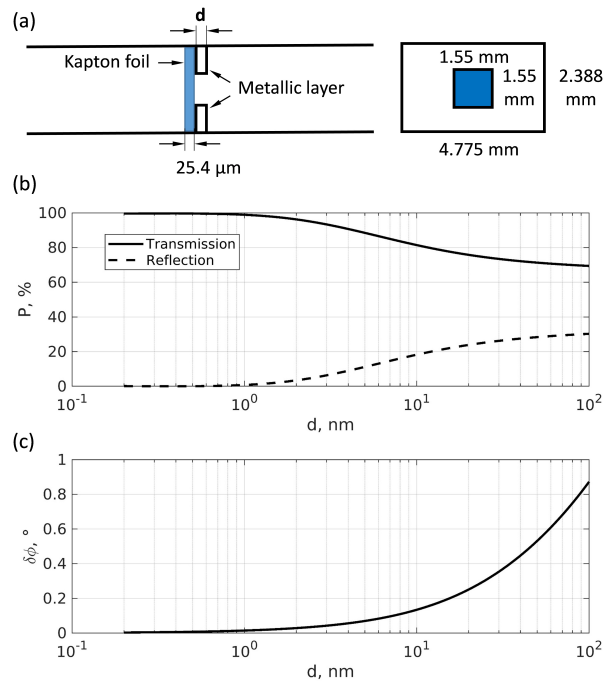

FIG. 15. Simulations of the effect of a thin metallic layer on the Kapton foil in the MILS waveguide on the power and phase measurements: (a) model setup, sizes in mm, (b) transmission and reflection power fractions and (c) phase shift as a function of the metallic layer thickness.

Up to 30 % of power transmission reduction can be achieved with the largest layer thickness (Fig. 15b). This range of values suffice to explain the power drift observed in the experiments, but unfortunately there are no measurements of the deposited metallic layer properties, which could prove or disprove this possible explanation. Therefore, in order to correct the possible power level drift, only the procedure explained in the previous subsection is used in the data post-processing and no correction is done for the suspected influence of a metallic layer, until further investigation on the metal deposition is done. It was also verified that the influence of the metallic layer on the phase measurement is negligible (Fig. 15c).

## D. Cross-talk between the detector output channels

At the beginning of MILS experimental setup commissioning, it was noticed that the output phase signal could vary depending on the ratio of the power of the two input signals, while the input phases on both channels are kept constant. In order to exclude the cross talk between the two channels, the circuit was modified to ensure 35 dB isolation between the channels. However, the observed phase variation was not significantly affected by this change.

The reason for the phase error might be either in the MILS detector or in the signal source. Another phase and power detector, Rohde & Schwarz ZPV Vector Voltmeter, was employed to check the variation of the phase. It was found that both the MILS detector and the ZPV Vector Voltmeter register a stepwise variation of the signal (Fig. 16). This variation is specific for each signal source, as it was checked for the signal generator Siglent SDG 6022X, used in the MILS setup at ASDEX Upgrade, and another signal generator Rigol 4162. Even though there are some deviations in the measured errors, the steps correlate between the two measurements methods, as well as coincide with the switching of the internal step attenuators of the signal generators, which can be heard during the calibration. It can be assumed that the MILS detector is not responsible for the observed error and the phase measurements with constant power output of the signal generator should be insensitive to the power differences in the two input channels. It is important, since the signal going through the plasma is expected to vary within several orders of magnitude.

From the MILS microwave electronics, the DC signals from the detector output need to be transferred to the data acquisition computer. The analog-to-digital converters (ADC), from which light fibers deliver the data to the computer, are located in a different sector of the AUG torus hall (tens of meters away). The signals are transferred by coaxial cables from one sector to the other. This setup can introduce noise and, moreover, the grounding in different sectors is independent, which can lead to parasitic signals. In order to avoid ground loops, isolation amplifiers (Analog Devices AD215BY) are used before the ADCs.

The analysis of the acquired signals showed a noise level of 3.5 mV RMS on the amplitude and phase signals, of which the estimated error of ADCs is 0.3 mV. In a FFT analysis, peaks at 430 and 860 kHz were found, which correspond to the fundamental and the first harmonic of the DC-DC converter switching frequency of the isolation amplifiers (Fig. 17). Adding a 100 kHz low pass filter at the output of the isolation amplifiers reduced the noise level to 1.7 mV RMS and weakened the 430 and 860 kHz components significantly (Fig. 17). The next step to reduce the noise and to be able to use the data without the 100 kHz temporal resolution limit is to set up the data acquisition right next to the MILS electronics.

Since one of objectives of the diagnostic is to provide data for ICRF studies and many diagnostics are affected by strong RF pick-up during the ICRF operation, it was checked that ICRF-induced noise in the MILS signals is negligible.

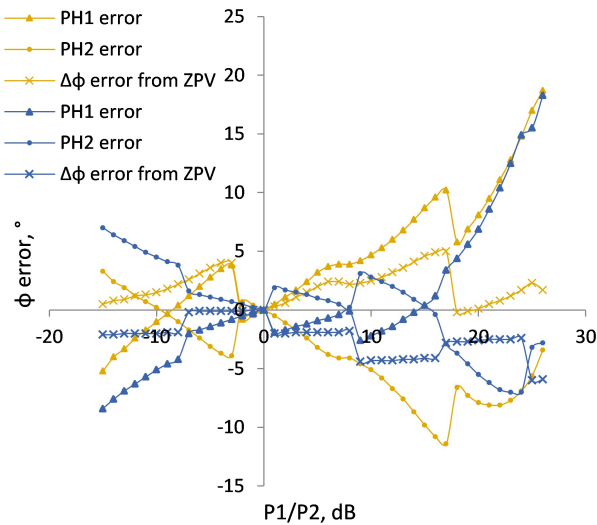

FIG. 16. Error in phase measurement depending on the ratio of the input power in the two channels. Colors correspond to different signal generators used for input, yellow – RIGOL, blue – SIGLENT. PH1 and PH2 signals are the outputs of the MILS detector and additionally the error in phase is measured by ZPV Vector Voltmeter.

### E. Ground loops and noise

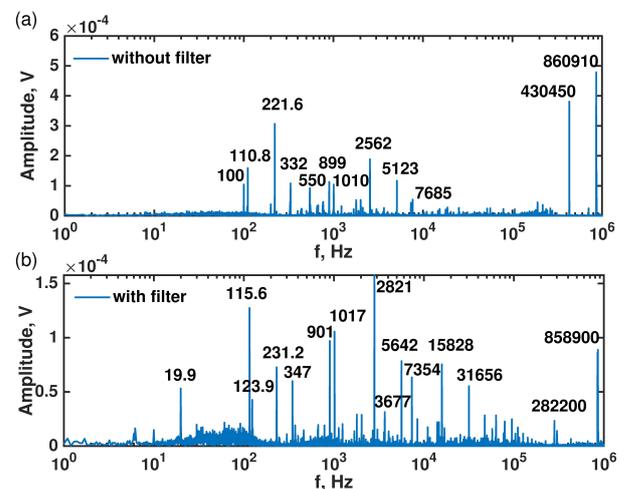

FIG. 17. FFT of the MILS signal: (a) without and (b) with 100 kHz low pass filter. Note the reduced maximum amplitude in the bottom plot.

### F. Microwave radiation from plasma

There is no significant radiation sources in ASDEX Upgrade typical plasmas at the frequency of MILS operation (47 GHz). The frequencies of Electron Cyclotron Resonance Heating (ECRH) are 140 and 105 GHz. The typical measured Electron Cyclotron Emission from the AUG plasma is in the range of 100–140 GHz. In addition, the direction of the MILS antennas tangentially to the plasma ensures good protection from any background radiation from most of the plasma volume. It was observed with MILS turned off, that the registered radiation is at least 25 dB lower than the MILS signal level.

### G. Summary on the sources of MILS measurements distortion

While the stray magnetic field and the in-vessel thermal loads can lead to a loss or strong distortion of the signal, other potential sources of noise or data corruption have been proven not to affect the MILS measurements.

It has become clear what components can potentially cause failures due to the magnetic field and strategies were developed and implemented to ensure their reliable operation. In order to compensate for the power and phase signals drift due to the radiated energy (or potentially due to metallic impurities deposition inside the MILS waveguide), a post-processing algorithm has been adapted.

The total error in the phase and power measurements comes from two of the factors described above, the noise of 1.7 mV, equal to constant errors $\delta\varphi_{noise} = 0.17°$ and $\delta P_{noise} = 0.8\,\%$, plus the error due to the correction for the phase and power drift $\delta\varphi_{drift}$ and $\delta P_{drift}$ (time dependent, because the radiated energy is time dependent). The latter one can dominate over all other errors, if the radiated energy is very large. In the studied examples, there is one such case, and in the others the experimental and the modelling errors are comparable.

## VI. CONCLUSIONS

The recently developed diagnostic technique for electron density measurements, intensity refractometry, has been experimentally tested in application to the tokamak edge. The comparison of intensity refractometry to other diagnostics in AUG experiments yields persistent agreement of the density values in front of the limiter, within one standard deviation range. In the limiter shadow, consistent density estimation is given by different diagnostics in many cases, but in some higher-density cases density values from different diagnostics disagree, lying outside of the interval of three standard deviations. The main cause for such discrepancies is expected to be the dissimilar local conditions in the limiter shadow for different diagnostics, including different toroidal proximity to the local limiter, various radial distance to the global limiter caused by imperfect flux surfaces alignment with the limiter, etc.

The comparisons highlight the presence of the poorly diagnosed region in the far SOL, which is now covered by MILS and for which data can be provided routinely, since the diagnostic is not perturbing and can be operated in any plasma conditions (not affected by factors like stray radiation or RF pick-up). The presented examples cover a large range of density values in the measured region, from $2*10^{17}$ m$^{-3}$ to $2*10^{19}$ m$^{-3}$ at the limiter position, while the accessible MILS range, explored in the theoretical study[8], is even wider. The radial range of MILS measurements can be observed, which depends on the density. In the given examples, the radial positions varied from 5 cm in front of the limiter (up to 1 cm inside the separatrix was measured) to 3 cm in the limiter shadow, with a span of the MILS density profile length from 2.5 to 6.5 cm. The high temporal resolution of MILS allows accurate separation of the background density and density fluctuations, so it was checked that the processed data does not contain any significant fluctuations, which could distort the results. Examples of such fluctuations have been previously identified in the MILS data[9]. A detailed analysis of the typical density oscillation level in the far SOL is out of the scope of the current study and is planned as a future work.

The experimental factors, which may affect the functioning of the diagnostic or disturb the data, have been analyzed and solutions for the various aspects implemented. In particular, the sensitivity of the electronics components to the stray magnetic field can lead to the power outage of the diagnostic and such components should either be replaced or monitored for failures. The in-vessel parts (waveguides) heating was identified as one of the largest sources of the errors in the data. The post-processing allows good compensation of the drifts in the measured quantities, but an uncertainty range is given to account for possible errors in the drift compensation. In the future, the issue of the signals drift has to be addressed by hardware improvement.

Techniques for data processing and density reconstruction, previously developed for MILS[8], prove to be applicable for processing experimental data. In most of the cases, values of the experimental errors are comparable to the level of the modelling errors. The total uncertainties are low enough to keep the numerical methods applicable and to have good accuracy of density reconstruction. The current numerical methods are versatile and flexible and allow easy incorporation of improvements. The obtained experimental and numerical results provide hints for the understanding of the physics of the MILS probing wave propagation, refraction in the plasma and interference at the receiver. This knowledge will be useful for future optimizations of the density reconstruction algorithm and of the MILS design parameters.

The previous theoretical study[8] and the newly obtained experimental validation of the applicability of the intensity refractometry principle in the conditions of the tokamak edge plasma open a range of possibilities of the further use of this novel diagnostic technique. In tokamaks, stellarators, linear devices, etc., whenever the tangential wave propagation can be used and suitable diagnostic parameters can be found, the principles of the developed measuring technique and data processing can be applied and further refined. As another possible application, plasma position control with the help of intensity refractometry could be attempted in the devices like DEMO. Since it can provide real-time measurements without the need to average over long periods and can target low densities at the edge, the position of the plasma outer boundary can be deduced and feedback-controlled from the level of the intensity change of the probing beam (or tomography with several beams). An applicability of the intensity refractometry to local magnetic field measurements could be considered as well, if the probing wave in the X-mode polarization is used and suitable conditions are found to achieve enough sensitivity for such measurements.

**ACKNOWLEDGEMENTS**

The authors thank G. Siegl and J. Kneidl for the diagnostic installation and maintenance in AUG, R. Ochoukov for the helpful assistance, M. Faitsch for the tests with infrared measurements and G. Conway for the consultation about microwave techniques.

This work has been carried out within the framework of the EUROfusion Consortium, funded by the European Union via the Euratom Research and Training Programme (Grant Agreement No 101052200 — EUROfusion). Views and opinions expressed are however those of the author(s) only and do not necessarily reflect those of the European Union or the European Commission. Neither the European Union nor the European Commission can be held responsible for them.

**DATA AVAILABILITY**

The data that support the findings of this study are available from the corresponding author upon reasonable request.

**APPENDIX**

**A. Reconstruction of density profile from two measured values**

The shape of the density profile, chosen for the MILS database construction, is described by three parameters. The diagnostic only provides two outputs, which can easily lead to the question whether such a profile can be reconstructed from just two quantities. However, an unambiguous mapping from a 3D space to a 2D space is possible, speaking strictly mathematically. A simple example is a line defined in the 3D parameter space and its projection on a surface. For any point of the line in the 3D space, a single corresponding point on the 2D line exists; the same is true the other way around. If the mapping is pre-defined, from the coordinates of any point on the 2D line, a unique set of the 3D coordinates on the 3D line can be reconstructed.

Examples of unambiguous mapping from 3D to 2D that are much more complex are possible as well. Therefore, the only question is whether we have a case of unambiguous mapping in our study. To find such profile parametrization, which allows an unambiguous mapping or as close as possible to unambiguous, would be an ideal case. The currently considered MILS setup and profile definition result in a case, which is not strictly unambiguous, but has important features, which allow partially using the unambiguous mapping and partially employing another method of the solution optimization. The main aspects of the MILS database are:

- There is a part of the parameter space, where the dependence of the $(\Delta\varphi, P/P_{vac})$ values on the parameters $(n_{lim}, a_{in}, a_{out})$ degenerates to only two parameters $(n_{lim}, a_{out})$, see Fig. 18. This is the region of high densities, for which the probing wave part reaching the receiver is refracted earlier than it can reach the plasma outside of the limiter shadow and therefore it is sensitive to the density in the limiter shadow only. In this part of the database, a mapping from 2D to 2D takes place, which is obviously unambiguous. The density range in this region is $n_{lim} \geq 3*10^{18}$ m$^{-3}$.
- A region of very low densities is present in the database, where the mapping between $(\Delta\varphi, P/P_{vac})$ and $(n_{lim}, a_{in}, a_{out})$ is very incoherent (Fig. 18). These densities are almost of no interest for practical applications and we do not focus on the density reconstruction in this part of the database. The approximate limit, below which the measurement point is located in this incoherent region, is $n_{lim} = 2*10^{17}$ m$^{-3}$.
- Except for these two regions, in the other part of the MILS database the position of a point in the $(\Delta\varphi, P/P_{vac})$ coordinates generally depends on all three parameters $(n_{lim}, a_{in}, a_{out})$. Consequently, points with different $(n_{lim}, a_{in}, a_{out})$ might be very close to each other on the $(\Delta\varphi, P/P_{vac})$ diagram. The important aspect here is that it does not happen completely randomly and there are clear patterns of how points group near each other on the $(\Delta\varphi, P/P_{vac})$ diagram. It allows suggesting a hypothesis that for density profiles of two points with close $(\Delta\varphi, P/P_{vac})$ values similar features can be found, which lead to the proximity of the $(\Delta\varphi, P/P_{vac})$ values. Since identifying these similar features in a simple analytical way does not seem to be possible, a numerical method is employed, which is suitable for complex problems – a genetic algorithm genMILS[8].

It is vital to recognize the existence of physical reasons (wave refraction and interference mechanisms in the chosen setup) behind the non-random points distribution on the $(\Delta\varphi, P/P_{vac})$ diagram. In an opposite case of points random placement, the diagnostic approach, including the employed numerical methods, would not be feasible.

By varying the MILS parameters or the way of the density profile parametrization, it is possible to achieve the mapping from the measured parameters to the parameters describing the density profile with lower or higher degree of unambiguity (well-chosen parametrization would allow higher degree of 3D parameter space degeneracy to 2D).

The work on such optimizations is ongoing. One of the insights into the question of what density profile features influence most the MILS output quantities is obtained from the currently employed genMILS algorithm.

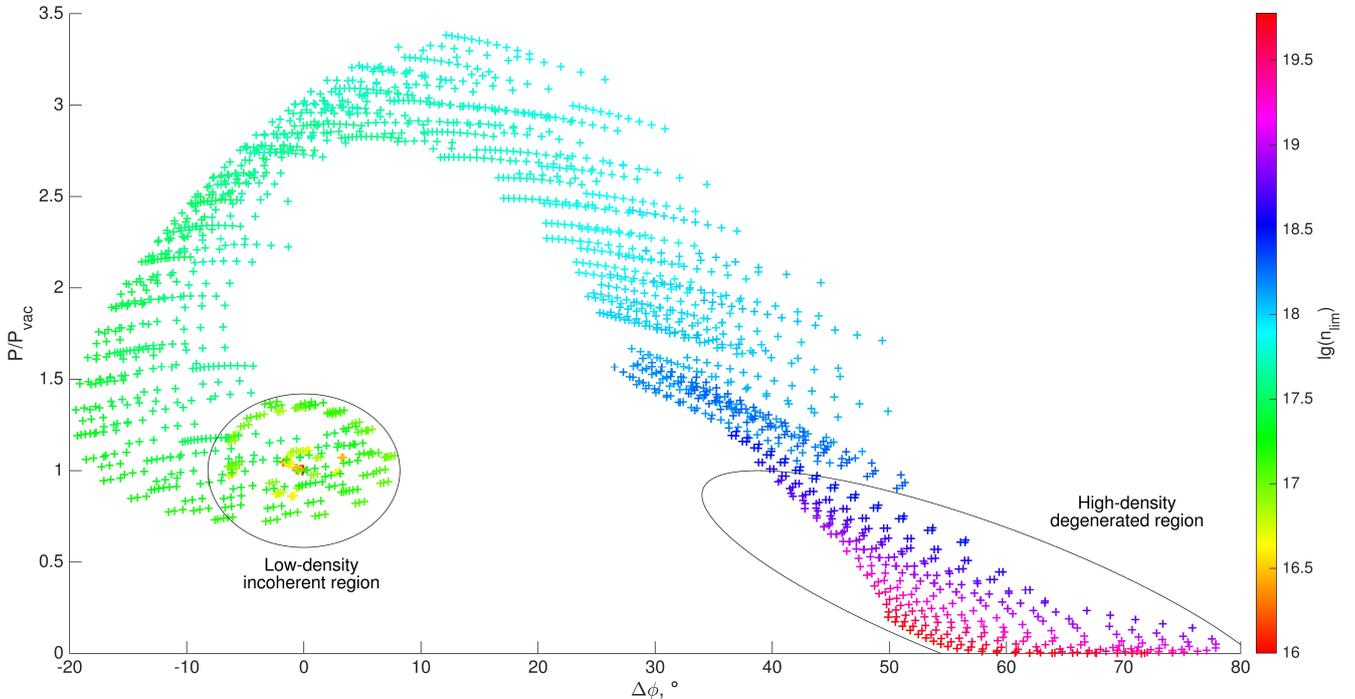

FIG. 18. MILS modelling database with parameter $n_{lim}$ shown by color. Two regions of the database with properties different from the main part of the database are indicated.

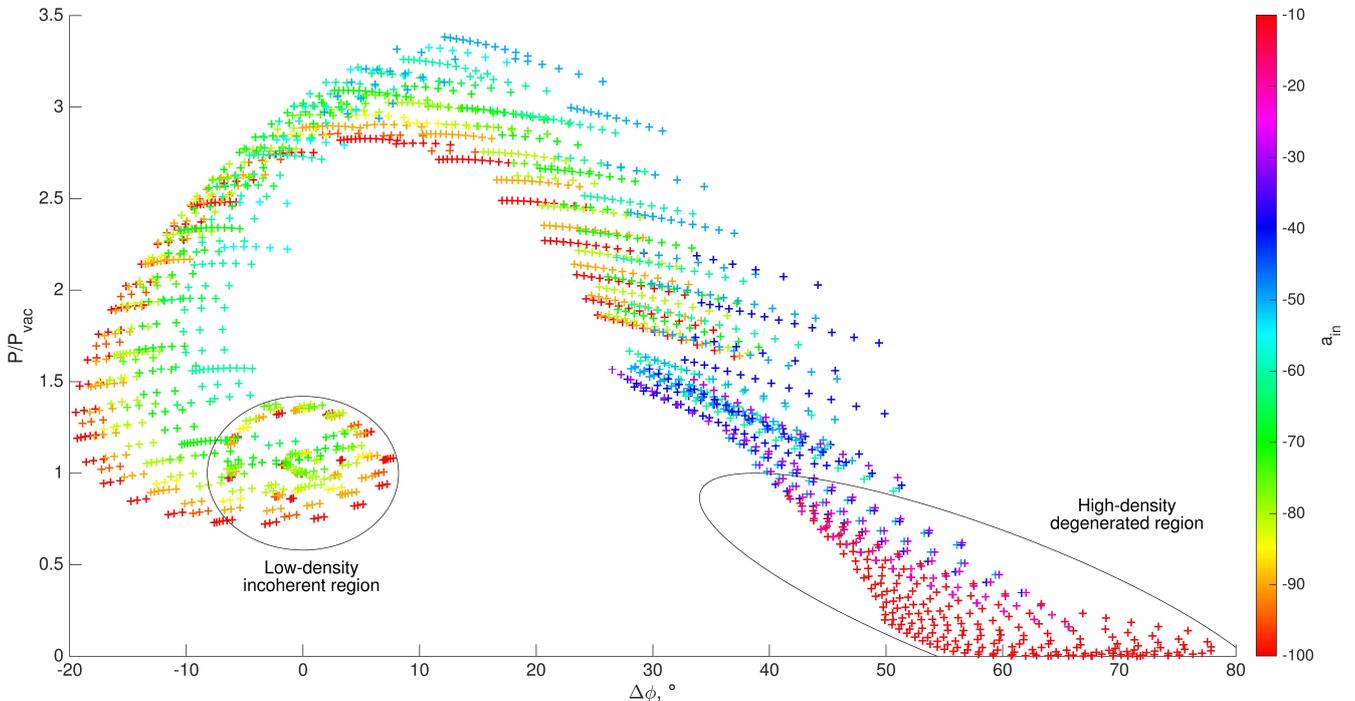

FIG. 19. MILS modelling database with parameter $a_{in}$ shown by color. Two regions of the database with properties different from the main part of the database are indicated.

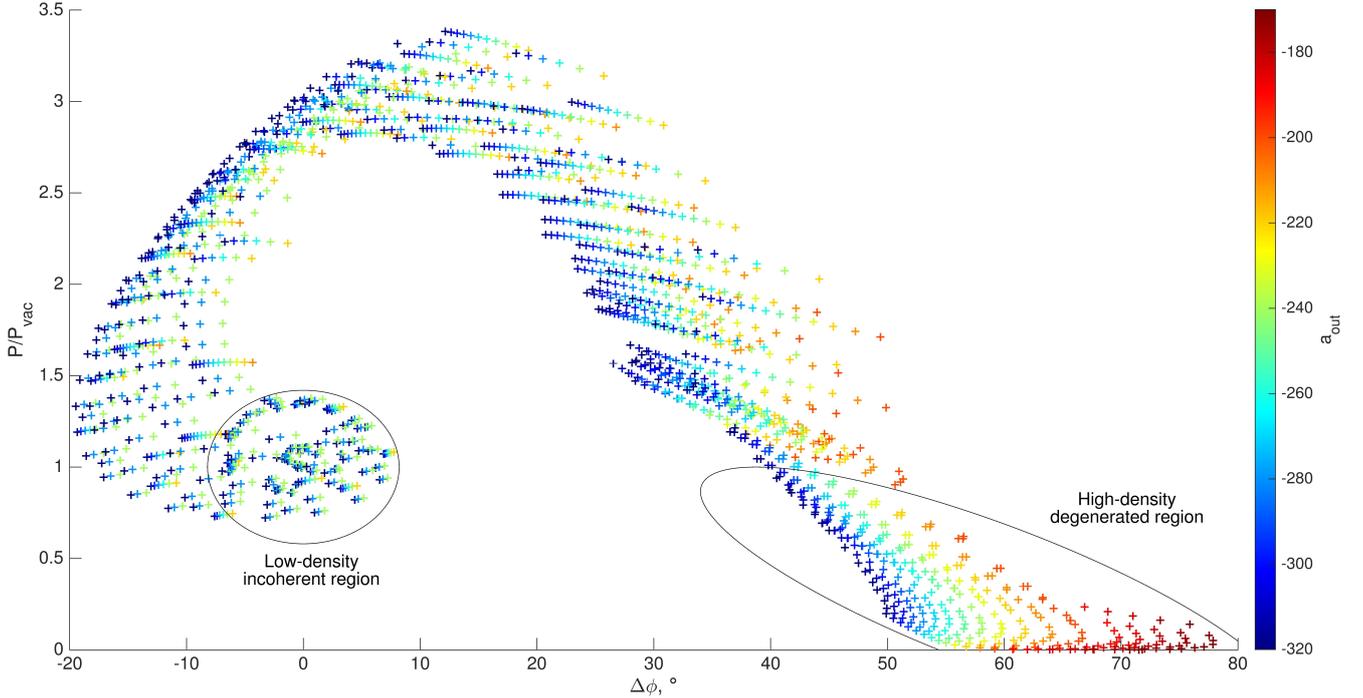

FIG. 20. MILS modelling database with parameter $a_{out}$ shown by color. Two regions of the database with properties different from the main part of the database are indicated.

**B. Genetic algorithm genMILS**

Genetic algorithms, being one of the Nature-inspired optimization algorithms, have been developed and applied for a wide range of optimization problems[18,19]. The idea of the genetic algorithms is to evolve a population of candidate problem solutions, performing a selection process at each generation to find the optimum and then further improving the genetic material by crossover and mutations. In our task, the set of the density profiles constitute the population and the "genes" chosen to describe each density profile are $m$ radially distributed points along the profile $n_m(r)$.

The classical genetic algorithm would start from defining $N$ sets of $n_m(r)$ as randomized values, which constitute the population of $N$ candidate solutions. In the selection process, the fitness function of each of the N solutions would be evaluated and the best candidate or the best combination of the candidates would be found. The next generation of the solutions population would depend on the fitness functions of the previous generation (choosing the "strongest parents" to carry on the best "genes") and would include some "mutation" as randomization. This process can be repeated until a stopping criterion is met.

In the current version of the genMILS algorithm, a simple version of this procedure is performed:
- Only one generation of candidate solutions is generated (the MILS database with double-exponential profiles) and the selection procedure is carried out once.
- The set of the initial $n_m(r)$ is not random; it is defined within the boundaries of the database, which sets a limit on the density minimum and maximum values, on the direction of the density decay (towards the tokamak wall) and on the profile shape (making the radial points to be not independent). While this reduction of randomization reduces the generality of the solution, it lowers the computational cost greatly and reduces the possible solutions to only those, which are considered relevant to our case of density profile at the edge of a tokamak.
- Since the selection procedure is only carried out once, we do not chose a single best candidate solution, but define a weighted combination of the candidate solutions from their fitness functions.

The developed algorithm remains quite generic in terms of its possible improvements. A different way of defining the original candidate density profiles can be implemented, another selection procedure can be chosen or more generations of selection and crossover/mutation processes can be added. One more advantage of the genetic algorithms is a possibility to easily introduce priors[19] into the computations. Priors could come, for example of our problem, from information about the density profile obtained from other diagnostics, and are introduced into the fitness function calculation. A limit on the density values at a certain location or on the profile steepness can be set and the addition of priors will ensure that the preferences is given to a certain kind of profiles during the selection. Another important aspect, common in the application of the genetic algorithms, is the question of the independency of the parameters ("genes"), which define the candidate solutions (radial points in the density profile in our case). A problem with correlated parameters could be transformed to a problem with independent parameters, if the covariance matrix of the parameters correlations can be defined[19]. In our case, there are too many parameters describing the candidate solution, which cannot be transformed to the same number of independent parameters, so the procedure, which is normally used for independent parameters, is used for the density profile points, which are dependent on their neighbours due to the fixed profile shape. Consequently, the found solution will consist of points, which should not be considered independent. Their correlation can be clearly seen[8], since the shape of the final profile remains very close to the double-exponential shape of the candidate solutions. The final solution of our optimization problem is defined as a weighted sum of candidate solutions. Generally, choosing a single best candidate instead or changing the way the weights are defined is possible. Our approach is somewhat similar to the one in the so-called Weighted Essentially Non-Oscillating (WENO) schemes for the flux-limited solutions of the hyperbolic equations[20], where a convex combination of stencil is used for the function reconstruction, instead of choosing only one having the best oscillation reducing score. The weights used in this combination are determined from the scores of individual stencils, with the higher weights prescribed to the solutions with the best oscillations suppression quality.

The exact definition of the choice of the $n_m(r)$ values, their total number $m$ and the fitness function (weight) for the selection procedure in the current version of the genMILS algorithm have been described in details and application examples given[8]. The density reconstruction for the data presented in this article was done using genMILS in the same way as described in the cited theoretical paper. The genetic algorithm is applied for the whole $(\Delta\varphi, P/P_{vac})$ parameter space, even though, as described above, two regions of the $(\Delta\varphi, P/P_{vac})$ diagram have their own specific features. Employing a single algorithm for any data processing of MILS is more efficient than developing specialized algorithms for the mentioned parts of the $(\Delta\varphi, P/P_{vac})$ parameter space. The consequence of it is the different level of the genMILS method error in the reconstructed profile (largest for the low-density region, smallest for the highest densities[8]).

The application of genMILS has made it clear that there is a certain characteristic of the reconstructed profile observed for all studied cases. The final solution is defined as a weighted sum of the densities from several candidate profiles, which have the highest values of the fitness function. Since each radial point of the final profile is calculated as independent, the spread of the density values in the candidate profiles varies radially (see error bars in Fig. 8 and Fig. 9 and examples in the theoretical study[8]). It is minimal close to the center of the profile and grows notably towards the profile edges. The chosen distance of 0.5 cm between the radial points turned out to be small enough to resolve these variations. In the essence, the central part (1-2 cm radially) of each database density profile (candidate profiles in genMILS) plays the dominant role in defining the output of the MILS diagnostic, i.e. the position of this database point on the $(\Delta\varphi, P/P_{vac})$ diagram. This observation allows better understanding of the process of MILS probing swave propagation and interference at the receiver, where the central (in terms of radial position) part of the received wave makes the dominant contribution and the details of the density distribution further away radially might not play any significant role for the measured values. This knowledge can be used for further optimization of the density profile parametrization and can guide future designs of new MILS configurations.